# Highly entangled photons from hybrid piezoelectric-semiconductor quantum dot devices

Rinaldo Trotta,[1]* Johannes S. Wildmann,[1] Eugenio Zallo,[2] Oliver G. Schmidt,[2] and Armando Rastelli[1]

[1]Institute of Semiconductor and Solid State Physics, Johannes Kepler University Linz, Altenbergerstr. 69, A-4040 Linz, Austria.

[2]Institute for Integrative Nanosciences, IFW Dresden, Helmholtzstr. 20, D-01069 Dresden, Germany.

Entanglement resources are key ingredients of future quantum technologies. If they could be efficiently integrated into a semiconductor platform a new generation of devices could be envisioned, whose quantum-mechanical functionalities are controlled via the mature semiconductor technology. Epitaxial quantum dots (QDs) embedded in diodes would embody such ideal quantum devices, but QD structural asymmetries lower dramatically the degree of entanglement of the sources and hamper severely their real exploitation in the foreseen applications. In this work, we overcome this hurdle using strain-tunable optoelectronic devices, where *any* QD can be tuned for the emission of highly polarization-entangled photons. The electrically-controlled sources violate Bell's inequalities without the need of spectral or temporal filtering and they feature the highest degree of entanglement ever reported for QDs, with concurrence as high as 0.75±0.02. These quantum-devices are at present the most promising candidates for the direct implementation of QD-based entanglement-resources in quantum information science and technology.

**Corresponding Author**

Dr. Rinaldo Trotta

Institute of Semiconductor and Solid State Physics,

Johannes Kepler University Linz, Altenbergerstr. 69,

A-4040 Linz, Austria.

Tel.: +43 732 2468 9599

Fax: +43 732 2468 8650

e-mail: rinaldo.trotta@jku.at

The potential of semiconductor QDs as entanglement resources was highlighted for the first time in 2000 by Benson and co-workers[1]. The proposal relies on the generation of polarization entangled photon pairs during the radiative decay of the biexciton (XX) to the exciton (X) to the crystal ground state (0) and it is analogous to the "atomic concept" used by Aspect and co-workers[2] to demonstrate the non-local nature of entanglement: In QDs featuring high structural symmetry ($D_{2d}$ or higher), the intermediate exciton level – built up from an electron and a heavy-hole – consists of two degenerate bright states with total spin projection ± 1. The XX (total spin projection 0) cascade to the ground state can thus take place following two different decay paths: (*i*) with the emission of a right-circularly-polarized biexciton photon ($R_{XX}$) followed by a left-circularly-polarized exciton photon ($L_X$) or (*ii*) with the emission of a $L_{XX}$ photon followed by a $R_X$ photon. Since the two decay paths have equal probability, the photon-pair results entangled in polarization and predicted to be in the maximally entangled Bell state

$\psi = \frac{1}{\sqrt{2}}(|R_{XX}L_X\rangle + |L_{XX}R_X\rangle)$, which can be equivalently rewritten as

$\psi = \frac{1}{\sqrt{2}}(|H_{XX}H_X\rangle + |V_{XX}V_X\rangle)$ or $\psi = \frac{1}{\sqrt{2}}(|D_{XX}D_X\rangle + |A_{XX}A_X\rangle)$, where *H* (*V*) and

*D* (*A*) indicate horizontally (vertically)-polarized and diagonally (antidiagonally)-polarized photons, respectively. This theoretical proposal has led to an explosion of research efforts[3-11], mainly motivated by the possibility of having at hand *solid-state triggered* entangled photon sources for quantum cryptography and quantum computation. However, the simplicity of the proposed scheme conceals the very stringent requirement set on the QD structural properties: the absence of an energetic splitting between the intermediate X states, commonly referred to as the fine structure splitting (FSS, *s*). The FSS – caused by mixing of the ± 1 spin states due to the anisotropic electron-hole exchange interaction[12] in QDs with reduced structural symmetry (lower than $D_{2d}$) – induces dephasing and dramatically affects the degree of entanglement. More precisely[13,14], the presence of a finite FSS turns the state into

$$\psi' = \frac{1}{\sqrt{2}}\left(|H_{XX}H_X\rangle + e^{i\frac{s\tau}{h}}|V_{XX}V_X\rangle\right),$$ where $\tau$ is the time interval between the emission of the two photons and $h$ is the Planck's constant. During the time $\tau$, the FSS causes the superposition of the photon-pair state to oscillate, thus reducing the time-averaged fidelity of $\psi'$ to the maximally entangled Bell state $\psi$, possibly revealing only classical correlations among the emitted photons[14]. Despite impressive progress in the fabrication of highly symmetric QDs[11] and in the post-growth control of their optical properties via external perturbations[15-18], it remains extremely difficult to systematically drive the FSS to vanishing small values, mainly due to the coherent coupling of the two bright exciton states[19]. As a result, the high degree of entanglement needed for the implementation of QDs in real applications and in advanced quantum optics experiments[20,21] – such as the one required to violate Bell's inequalities[22] – has been achieved only by few groups worldwide by cherry-picking the very rare QDs featuring $s$~$0$[23] or by using temporal[9,24] and spectral filtering[4] techniques, which inevitably reduce the brightness of the source. In this work, we demonstrate that virtually *any* QD embedded in hybrid semiconductor-piezoelectric devices can be used for the generation of *highly* polarization-entangled photons. This novel class of electrically-controlled entangled-photon sources violates Bell's inequalities without the need of temporal or spectral filtering. By means of quantum state tomography we prove that our source features the highest degree of

entanglement ever reported for QD-based entanglement resources, with a concurrence as high as $0.75 \pm 0.02$. This result is due to the unique capability of the electro-elastic fields generated by our device to suppress the FSS. Finally, we clarify that although moderately small FSS ($s < 4$ µeV) are sufficient to overcome the classical limit, a much tighter control over the FSS is required to violate Bell's inequalities ($s < 1$ µeV), which also sets the range of X energies (~1 meV) over which our device operates as energy-tunable source of highly entangled photons.

**RESULTS**

**Electro-elastic fields: erasing the fine structure splitting.**

The key feature of our device is the capability to provide two independent external perturbations for engineering the energy levels of the quantum emitters on demand. This is achieved by embedding InGaAs QDs in the intrinsic region of n-i-p diode-like nanomembranes and by their subsequent integration onto $[Pb(Mg_{1/3}Nb_{2/3})O_3]_{0.72}$-$[PbTiO_3]_{0.28}$ (PMN-PT) piezoelectric actuators[25]. Therefore, the device features two electrically-controlled tuning knobs (see Fig. 1a): the voltage applied to the diode allows the electric field ($F_d$) across the QDs to be controlled. Simultaneously, the electric field ($F_p$) across the piezoelectric actuator induces in-plane anisotropic biaxial strain fields to the QDs. Further details can be found in the methods.

In order to obtain high-fidelity entangled photons from the XX-X cascade, the coherent coupling of the bright exciton states has to be suppressed. For this purpose, it is fundamental to achieve full control over *two* QD parameters: the FSS *and* the polarization direction of the exciton emission ($\theta$). Controlling only one of them results inevitably in an anticrossing of the two X states as the external field is varied[15,17,19]. In the following, we explain the two-step procedure that allows the X-level degeneracy to be restored in *any* QD. (1) Strain ($F_p$) is first used to precisely align the polarization direction of the exciton emission ($\theta$) along the "effective" axis of application of the electric field ($F_d$), see inset of Fig. 1c. In the case of electric fields applied along the [001] crystal direction, this effective axis coincides with the [110] or [1-10][17]. For the particular QD reported in Fig. 1, the amount of strain needed to properly rotate $\theta$ is achieved at $F^*_p$= 17.3 kV/cm. (2) In spite the first step resulted in an increase of the FSS (from 8 μeV to 11 μeV, see the inset of Fig. 1c), the electric field is now capable to compensate completely for in-plane asymmetries in the QD confinement potential, and can thus drive the FSS through zero. This is shown in Fig. 1b-c, where the behavior of *s* and $\theta$ are reported as the absolute value of $F_d$ is ramped up: The FSS first decreases (linearly), it reaches the minimum value of *s* = (0.2 ± 0.3) μeV at $F^*_d$= -93 kV/cm and then it increases (linearly) again. On the other hand, $\theta$ remains constant and aligned along the [110] direction until an abrupt anti-clockwise rotation of 90 degrees takes place around the $F^*_d$ value at which the minimum FSS occurs. No appreciable changes of $\theta$ can be noticed as the magnitude of $F_d$ is further increased. The behavior of *s* and $\theta$ follow closely the theoretical predictions of a **k·p** model developed for describing the field-induced cancellation of the coherent coupling of the two bright exciton states, see the blue solid lines in Fig. 1b-c and reference 17.

Several important points need to be discussed at this stage: (*i*) The behavior of *s* and $\theta$ reported in Fig. 1b-c is *universal*, and always reproduced by the theoretical model. (*ii*) The values of $F_d$ and $F_p$ at which the FSS reaches *s*=0, *i.e.* $F^*_p$ and $F^*_d$, depend on the particular QD under study (see Fig. 3). This is connected to fluctuations in the composition, size, and shape among the different QDs that arise inevitably during their stochastic growth processes[26]. (*iii*) In order to restore the X level degeneracy, $F^*_p$ and $F^*_d$ have to fall within the tuning range, *i.e.*, for a fixed sample the probability *P* of finding QDs that can be driven to *s*=0 is determined by the tuning range. We found that *P* can be as high as 100% and it varies among different devices (the discussion of the differences among the devices is beyond the scope of the present work and will be presented elsewhere). This represents a remarkable improvement compared to previous works[4,5-11,15] and it allows us to correct *systematically* for the asymmetries in the confining potential that usually prevent QDs from emitting polarization-entangled photons, as we demonstrate in the following.

**Entangled photons at zero fine structure splitting.**

We now use polarization-resolved photon correlation spectroscopy to investigate in more detail the dynamics of the XX-X cascade in QDs whose FSS has been fine-tuned to *s*=0. The micro-photoluminescence spectrum of a typical QD acquired in these conditions is shown in Fig. 1d. The presence of the negatively (positively) charged exciton $X^-$ ($X^+$) appearing at the low (high) energy side of the X-XX doublet is very common in our QDs[27]. The inset of Fig. 1d displays the intensity of the exciton emission as a function of the angle the linear polarization analyzer forms with the [110] direction (see methods): it proves that the source is un-polarized (within 4%) at *s*=0. The first clear signature of entangled photon emission is shown in Figure 2a-f, where cross-correlation measurements between XX and X are reported for circular (Fig. 2a-d), linear (Fig. 2b-e) and diagonal (Fig. 2c-f) polarization basis for one of the investigated QDs. We observe strong bunching when recording coincidence counts with the following polarizations: $H_{XX} H_X$ (or $V_{XX} V_X$), $D_{XX} D_X$ (or $A_{XX} A_X$) and $R_{XX} L_X$ (or $L_{XX} R_X$). On the other hand, the bunching peaks disappear for $H_{XX} V_X$ (or $V_{XX} H_X$), $D_{XX} A_X$ (or $A_{XX} D_X$) and $R_{XX} R_X$ (or $L_{XX} L_X$). This is exactly the predicted behavior of photon pairs emitted in the maximally entangled Bell state $\psi$, where strong correlations for co-linear, co-diagonal, and cross-circular polarization are expected[3]. For an ideal source of entangled photons the degree of correlation (or correlation visibility) $C_{AB}$, defined as the difference between co-polarized and cross-polarized correlations divided by their sum, should be $C_{HV}=|C_{RL}|=C_{DA}=1$. In our experiment, we find (see Fig. 2g-i): $C_{HV}= 0.72\pm0.05$, $|C_{RL}|= 0.82\pm0.02$, $C_{DA}=0.72\pm0.05$. Although these values suggest that the photons emitted during the XX-X cascade are entangled, the deviations from the expected values indicate that the two-photon state is not exactly the Bell state $\psi$. We estimate the fidelity to $\psi$ via the following formula[13] $f = (1+ C_{HV} + C_{DA} + |C_{RL}|)/4$ and obtain $f = 0.82\pm0.04$ at zero time delay (see Fig. 2l). This value is well above the classical limit of 0.5 and highlights that the application of large external fields does not prevent QDs from emitting entangled photons. In order to completely characterize the entangled state, we have reconstructed the two-photon density matrix $\hat{\rho}$ measuring XX-X correlations in 36 polarization settings and with the help of a maximum likelihood method[28]. The real and imaginary parts of $\hat{\rho}$ for a selected QD are displayed in Fig. 3a-b, respectively. The presence of the strong off-diagonal terms $\hat{\rho}_{HH,VV}$ and $\hat{\rho}_{VV,HH}$ illustrates immediately the superposition of the two-photon pair wavefunctions. The matrix satisfies the Peres inseparability criterion[29] for entanglement, being -0.35 (<0) the minimum eigenvalue of its partial transpose. $\hat{\rho}$ contains also imaginary components (see Fig. 3b), which point out to the presence of a phase delay between $|H_{XX}H_X\rangle$ and $|V_{XX}V_X\rangle$. Therefore, the state is not exactly the maximally entangled Bell state $\psi$ but rather

$$\psi^* \approx \frac{1}{\sqrt{2}}\left(|H_{XX}H_X\rangle + e^{-i(0.23\pi)}|V_{XX}V_X\rangle\right),$$ which corresponds to the largest eigenvalue

$\lambda$=0.86 of $\hat{\rho}$. This phase delay arises most probably from a reflection at our beam splitter (see the Supplemental Material). The latter analysis confirms that the fidelity $f$ to the $\psi$ state is not the most appropriate parameter to quantify the degree of entanglement of our source and, following the work of James and co-workers[28], we extract from $\hat{\rho}$ of all the investigated QDs the following metrics: the tangle ($T$), concurrence ($C$) and entanglement of formation ($E_F$). For a perfect source of entangled (classical) light all these parameters should be equal to 1 (0). For the best QD studied here we obtain $T$=0.56 ± 0.03, $C$=0.75 ± 0.02, $E_F$=0.66 ± 0.05. These values rate as the highest ever reported for QD-based photon sources [4-11,13-15,18,23,24] and, most importantly, for an electrically-controlled optoelectronic device[9,10,15] without the aid of temporal or spectral filtering. The level of entanglement achieved is not yet ideal and mainly limited by depolarization of the X states caused by fluctuating QD environment[14] and recapture processes[8] (for a detailed discussion see the Supplemental Material). Temporal post-selection of the emitted photons[8,9,24] have been successfully employed to alleviate these effects and also residual non-zero FSS, though at expense of the brightness of the quantum source. We have investigated this strategy on our QDs and we have found an increase of all the parameters quantifying entanglement, with a concurrence as high as 0.82 when 60% of the XX-X coincidences are discarded (for more details see the Supplemental Materials). We believe that the level of entanglement of our source can be further improved using quasi-resonant excitation techniques, faster photons detectors and optimized growth protocols.

The importance of our results is not limited to the high entanglement-degree but it is the unique possibility of achieving this result virtually in all the QDs in the device. This is summarized in Fig. 3c-d, where we report the fidelity to $\psi$, the largest eigenvalue ($\lambda$) and the concurrence estimated from $\hat{\rho}$ for all the 3 QDs investigated in this work. Despite the QDs differ in terms of emission energy, linewidth, XX binding energies, $F^*_p$ and $F^*_d$ values, they all show a very high level of entanglement, with *average* concurrence (largest eigenvalue $\lambda$) equal to 0.72 (0.86). It is rather difficult to identify the origin of the small fluctuations clearly visible in Fig. 3c-d, since they could arise from the sources of entanglement degradation discussed above. In spite of these small differences, our results are in stark contrast to the large spread of values that can be found in the literature. We ascribe this spread to the lack of control over the FSS characterizing previous works, *i.e.*, we finally prove that an efficient and reproducible method to generate polarization entangled photons via the XX-X cascade exists.

**Electrically-controlled non-locality: violation of Bell's inequalities**

We now investigate how the fidelity *f* to the maximally entangled Bell state $\psi$ changes when we drive *s* through zero. This is shown in Fig. 4a for one of the investigated QD, where we let $F_d$ vary the splitting at $F^*_p$= 17.3 kV/cm. For the sake of clarity, the FSS is displayed with negative values for $F_d > F^*_d$. The fidelity first increases, it reaches the highly non-classical value of 0.82±0.04 and then it decreases again. As shown by Hudson and co-workers[13], this behavior can be approximated by a Lorentzian function, which – in our case – has a full-width at half maximum (FWHM) equal to 3 µeV. This value nicely matches the lifetime of the exciton transition ($\tau \sim 1$ ns, being roughly constant for the range of electric fields explored during the experiment), thus proving that significant entanglement can be measured once the FSS is reduced below the radiative linewidth of the X transition. For the particular QD under study, violation of the classical limit (see dashed line in Fig. 4a) can be observed even for *s* values as large as ~ 4 µeV. These results confirm the general idea that entanglement persists despite a non-zero separation between the two bright excitonic states[13], and explain the large spread of *f* values that can be found in the literature. However, overcoming the classical limit is not sufficient for applications relying on non-local correlations between the emitted photons, such as entanglement swapping[20] (the key element of quantum relays and repeaters) as well as quantum cryptography protocols[22]. One possible criterion to define the "useful" entanglement degree is the violation of Bell's inequalities, initially proposed to demonstrate the entanglement non-locality and then used as a base for quantum cryptography[22]. Following references 9 and 24 we use three different Bell's inequalities for three different planes of the Poincaré sphere: $S_{RC} = \sqrt{2}(C_{HV} - C_{RL})$, $S_{RD} = \sqrt{2}(C_{HV} + C_{DA})$, $S_{CD} = \sqrt{2}(C_{DA} - C_{RL})$. A value larger than 2 of one of these parameters ensures violation of Bell inequalities, even if our state deviates from the maximally entangled Bell state (see above). Fig. 4b shows the evolution of these parameters as a function of *s*. Paralleling the fidelity, a Lorentzian-like behavior is observed. The different FWHMs reflect the different values of the correlations in the linear, diagonal, and circular bases, see Fig. 2. Around the minimum splitting we measure $S_{RD} = 2.04 \pm 0.05$, $S_{CD} = 2.24 \pm 0.07$, $S_{RC} = 2.33 \pm 0.04$. The latter parameter shows violation of Bell's inequalities by more than 8 standard deviations and proves unambiguously that our electrically-driven source can produce non-local states of light. Finally, Fig. 4b highlights that violation of all Bell's inequalities can be achieved only for a very small range of FSS, $s < 1$ µeV. It is important to notice that we achieve these results using the raw data, without any temporal[8,9,24] or spectral[4] filtering and without any background light subtraction. The values of the three Bell parameters further increase well above the limit of 2 using temporal post-selection of the emitted photons, with a maximum value of $S_{RD} = 2.22 \pm 0.05$, $S_{CD} = 2.50 \pm 0.07$, $S_{RC} = 2.43 \pm 0.04$ (see the Supplemental Material). Having achieved such a high level of entanglement, it is worth discussing the possibility to use our device as *electrically-controlled energy-tunable source of entangled photons*, *i.e.*, a device where the energy of the X (or XX) transition is modified electrically while maintaining a high level entanglement. This is indeed a crucial prerequisite for entanglement swapping experiments using dissimilar QD-based qubits, and cannot rely on as-grown QDs only, *i.e.*, without the aid of external perturbations. Fig 4a-b show that violation of the classical limit can be easily achieved when the X energy is tuned over a spectral range as large as ~ 4 meV. However, highly entangled photons can be emitted only in a narrower spectral range (0.6 meV), which can be further increased using temporal filtering to 1 meV (see the Supplemental Materials). Despite these values could be sufficient for proof of principle experiments aiming at interfacing distant QD-based entanglement resources, a different device concept is required to further increase the tunabilty of the X energy. We leave this point to future studies.

**DISCUSSION**

We have demonstrated that virtually *any* semiconductor QD embedded into strain-tunable optoelectronic devices can be used for the generation of entangled photons featuring the highest degree of entanglement reported to date for QD-based photon sources, with concurrence as high as 0.75 ± 0.02. The use of moderate temporal filtering further increases the concurrence to 0.82 at the expense of a 60% reduction of count rate. We have achieved these results by merging the semiconductor and piezoelectric technologies, which allow the long standing problem plaguing optically active quantum dots – the presence of an energetic splitting between the two bright exciton states – to be systematically solved. Furthermore, we prove that this novel class of quantum-devices violates Bell's inequalities without the need of temporal or spectral post-selection of the photons and can be used as electrically-controlled energy-tunable source of entangled photons, where highly polarization-entangled photons can be emitted over a ~ 1 meV spectral range of the exciton (or biexciton) energies.

After more than a decade from the first proposal[1], we have demonstrated that the implementation of QD-based optoelectronic devices into a solid-state quantum technology is feasible and we envisage that our device concept is going to play a key role in future experiments and applications built up around entanglement non-locality. The usefulness of the method is not limited to the InGaAs QDs investigated here. In particular, its implementation on strain-free GaAs QDs[10,23,30] may yield even larger entanglement levels due to the reduced nuclear magnetic fields and associated fluctuations.

**METHODS**

**Sample growth and device fabrication.** In(Ga)As QDs were grown by molecular beam epitaxy. Following oxide desorption and buffer growth, a 100 nm thick $Al_{0.75}Ga_{0.25}As$ sacrificial layer was deposited before the following $GaAs/Al_{0.4}Ga_{0.6}$ layers: a 180 nm thick n-doped layer, a 150 nm thick intrinsic region containing the QDs, and a 100 nm thick p-doped layer. The QDs were grown at 500°C and capped by an indium flush technique. The sacrificial layer in combination with metal evaporation, optical lithography and wet-chemical etching were used to release 500 nm thick nanomembranes of rectangular shape (150x120 $\mu m^2$). Gold thermo-compression bonding was then used to transfer the nanomembranes onto 300 μm thick PMN-PT actuator and 25 μm aluminum wires were used to connect electrically the device to a chip carrier. Further details on the device fabrication and performances can be found elsewhere[25].

**Micro-photoluminescence and photon-correlation spectroscopy.** Conventional micro-photoluminescence spectroscopy was used for the optical characterization of the devices. The measurements were performed at low temperature (typically 4-10 K) in a helium flow cryostat. The QDs were excited non-resonantly at 850 nm with a femtosecond Ti:Sapphire laser having an 80 MHz repetition rate and focused by a microscope objective with 0.42 numerical aperture. The same objective was used for the collection of the photoluminescence signal, which was spectrally analyzed by single or double spectrometers featuring 0.75 m focal length per stage and equipped with 1200 or 1800 lines/mm gratings, and finally detected by a nitrogen-cooled silicon charge-coupled device. Polarization-resolved micro-photoluminescence experiments were performed combining a rotating half-wave plate and a linear polarizer placed before the entrance slit of the spectrometer. The transmission axis of the polarizer was set parallel to the [110] direction of the GaAs crystal (within 3°) and perpendicular to the entrance slit of the spectrometer, which defines the laboratory reference for vertical polarization. The FSS and the polarization angle of the excitonic emission were evaluated using the same procedure reported in Ref. 24, which ensures sub-microelectronvolt resolution.

For photon-correlation measurements, the signal was split into two parts after the microscope objective using a non-polarizing 50/50 beam splitter, spectrally filtered with two independent spectrometers tuned to the XX and X energies (the band-pass window of the spectrometers, ~ 100 μeV, is much larger than the typical linewidth of both transitions), and finally sent to two Hanbury Brown and Twiss setups (HBT) at the exits of the spectrometers. Each HBT consists of a polarizing 50/50 beam splitter placed in front of two avalanche photodiodes (APDs), whose output is connected to a 4-channel correlation electronics for reconstructing the second-order cross-correlation function between the XX and X photons. The temporal resolution of the system is about 400 ps, mainly limited by the time jitter of the APDs. In order to select the appropriate polarization basis for cross-correlation measurements, properly oriented half-wave plates and quarter-wave plates were placed right after the first non-polarizing beamsplitter. The experimental setup allows the second order cross-correlation function $g_{AB}^{(2)}$ in 4 different polarization settings (*AB*) to be evaluated with a single measurement, *i.e.*, $g_{AB}^{(2)}$, $g_{AA}^{(2)}$, $g_{BB}^{(2)}$, and $g_{BA}^{(2)}$ were measured simultaneously. This reduces considerably the time of the experiment and, consequently, the effect of possible sample drifts during correlation measurements.

**Data analysis for correlation measurements.** Raw data were used in the analysis, without any background light subtraction. The second order correlation function was evaluated using the following formula: $g^{(2)}_{AB} = R/(N/n)$, where $R$ is the number of pairs detected in a 10 ns window centered at zero time-delay, $N$ is the number of pairs detected in the side-peaks and $n$ is the number of side-peaks considered (20 in our case, each of them integrated over 10 ns). For a given polarization base AB, the degree of correlation was calculated as explained in the text and averaging the two possible polarization combinations, *i.e.* $C_{AB} = \frac{1}{2}\left(\frac{g^{(2)}_{AA} - g^{(2)}_{AB}}{g^{(2)}_{AA} + g^{(2)}_{AB}} + \frac{g^{(2)}_{BB} - g^{(2)}_{BA}}{g^{(2)}_{BB} + g^{(2)}_{BA}}\right)$. For the density matrix reconstruction we have used probabilities, calculated from the raw counts via the following formula: $P_{AB} = \frac{g^{(2)}_{AB}}{g^{(2)}_{AA} + g^{(2)}_{AB} + g^{(2)}_{BA} + g^{(2)}_{BB}}$. The errors for all the quantities used in the analysis, including fidelity and the three Bell's parameteres listed in the main text, were propagated assuming a Poissonian distribution for $R$ and $N$ (and no error for $n$), *i.e.*, $\Delta R = \sqrt{R}$ and $\Delta N = \sqrt{N}$. It is worth mentioning that using raw counts (integrated counts of the zero-time delay peak) instead of probabilities leads to very similar results, with maximum concurrence of C=0.76±0.02.

ACKNOWLEDGMENT

We thank G. Bauer and A. Predojevic for fruitful discussions, P. Atkinson for help with the sample design and growth and F. Binder, A. Halilovic, U. Kainz, E. Vorhauer, and S. Brauer for technical assistance. The work was supported financially by the European Union Seventh Framework Programme 209 (FP7/2007-2013) under Grant Agreement No. 601126 210 (HANAS).

AUTHOR CONTRIBUTIONS

R.T. and A.R. conceived and designed the experiment. R.T. and J.S.W. fabricated the device, performed measurements and data analysis with help from A.R.. E.Z. and O.G.S carried out sample growth. R.T. wrote the manuscript with help from all the authors.

COMPETING FINANCIAL INTERESTS: The authors declare no competing financial interest.

**Figure 1. Suppression of the exciton fine structure splitting via electro-elastic fields**. (a). Sketch of InGaAs semiconductor quantum dots (QDs) embedded in a dual-knob device where in-plane anisotropic biaxial strain – induced by the electric fields across the piezoelectric actuator ($F_p$) – and vertical electric fields across the diode ($F_d$) are used to correct for the QD structural asymmetries that prevent them from emitting polarization entangled photons (shown by the two photons connected via a chain). (b). Behavior of the fine structure splitting (FSS, $s$) as a function of $F_d$ for a QD whose polarization direction of the exciton emission ($\theta$) has been previously aligned via $F_p$ along the effective direction of application of the electric field. The blue line corresponds to the prediction of the theoretical model, see Ref. 17. (c). Same as (b) for $\theta$. The inset shows the dependence (in polar coordinates) of $\Delta E$ as a function of the angle the linear polarization analyzer forms with the [110] crystal axis (polarization angle), where $\Delta E$ is half of the difference between XX and X energies minus its minimum value (see Ref. 17). The length and orientation of the petals give the value of $s$ and $\theta$, respectively. Blue and red circles are obtained respectively at ($F_d$, $F_p$) = (-30 kV/cm, 0 kV/cm) and ($F_d$, $F_p$) = (-30 kV/cm, 17.3 kV/cm), The latter value of $F_p$ (=$F^*_p$) corresponds to the critical strain that allows the bright exciton level degeneracy to be restored via $F_d$. For the blue (red) curve $s = 8$ µeV ($s = 11$ µeV) (d). Micro-photoluminescence spectrum of a representative QD at $s$=0. Radiative recombinations ascribed to the exciton (X), biexciton (XX) positively ($X^+$) and negatively ($X^-$) charged excitons are also indicated. The inset shows the intensity (in arb. units) of the exciton emission as a function of the polarization angle. 0° correspond to the [110] direction of the GaAs crystal.

**Figure 2. High-fidelity entangled photons from strain-tunable devices. (a)-(f)**. Polarization resolved cross-correlation measurement of exciton and biexciton photons emitted by a QD whose fine structure splitting has been fine-tuned to zero. $R_{XX,X}$ ($L_{XX,X}$), $D_{XX,X}$ ($A_{XX,X}$), $H_{XX,X}$ ($V_{XX,X}$) indicate respectively right (left) circular polarized photons, diagonally (anti-diagonally) polarized photons, and horizontally (vertically) linearly polarized photons. The different panels show the raw coincidence counts, without normalization (see methods). **(g)-(i).** Degree of correlation (see methods) in the diagonal (g), linear (h) and circular (i) bases for the same QD as in (a)-(f). **(l)**. Fidelity to the maximally entangled Bell state for same QD as (a)-(i). The dashed line indicates the classical limit (0.5).

**Figure 3. Quantum state tomography at zero fine structure splitting. (a)-(b)**. Real (a) and imaginary (b) part of the two-photon density matrix reconstructed via quantum state tomography in a QD at $s=0$. The matrix has been calculated using X-XX cross-correlation measurements in 36 polarization settings and with the aid of a maximum likelihood method. **(c)**. Bar chart of the fidelity $f$ to $\psi$ (red) and largest eigenvalue $\lambda$ (blue) for 3 different QDs fine-tuned to $s=0$. **(d)**. Bar chart of the concurrence for the same QDs whose fidelity values are shown in (c). The values of $F^*_p$ and $F^*_d$ at which $s\sim 0$ are also indicated.

**Figure 4. Violation of classical and Bell limits. (a)**. Fidelity as a function of the FSS in a QD at $F_p= 17.3$ kV/cm. The FSS is controlled via the electric field on the diode ($F_d$). The red solid line is the result of a Lorentzian fit to the data. The dashed line indicates the classical limit (0.5), while the range of X energy over which non-classical correlations between the emitted photons are measured is indicated by a solid bar. **(b)**. Bell parameters as a function of the FSS as in (a). The red, black, and blue full points indicate the $S_{RC}$, $S_{RD}$, $S_{CD}$ parameters, respectively. The solid lines are Lorentzian fits to the data. The dashed line indicates the Bell limit (2), while the range of X energy over which non-local correlations between the emitted photons are measured is indicated by a solid bar.

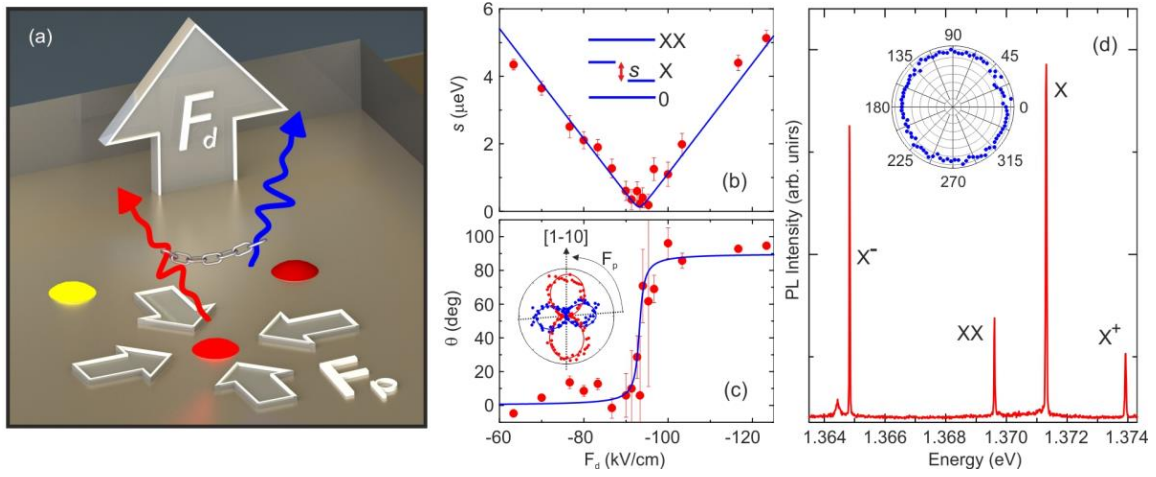

Figure 1 of 4
by R. Trotta *et al*.

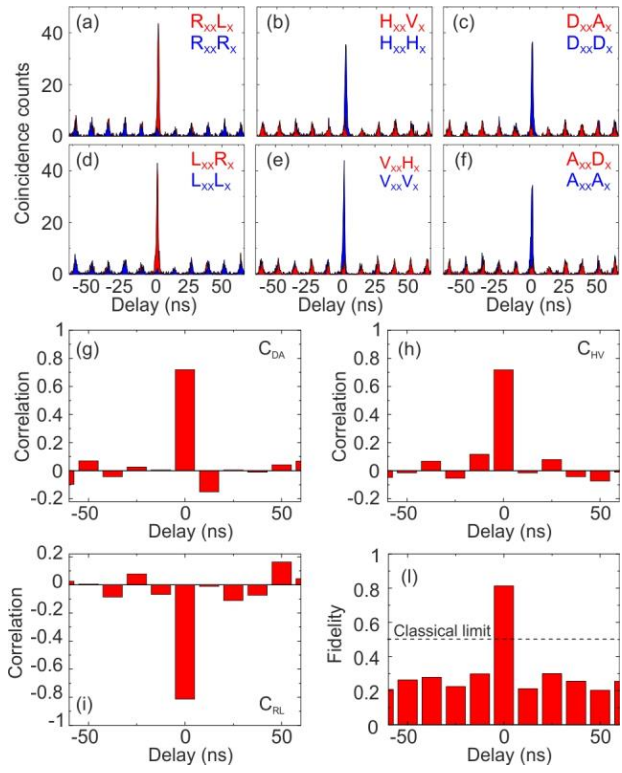

Figure 2 of 4
by R. Trotta *et al*.

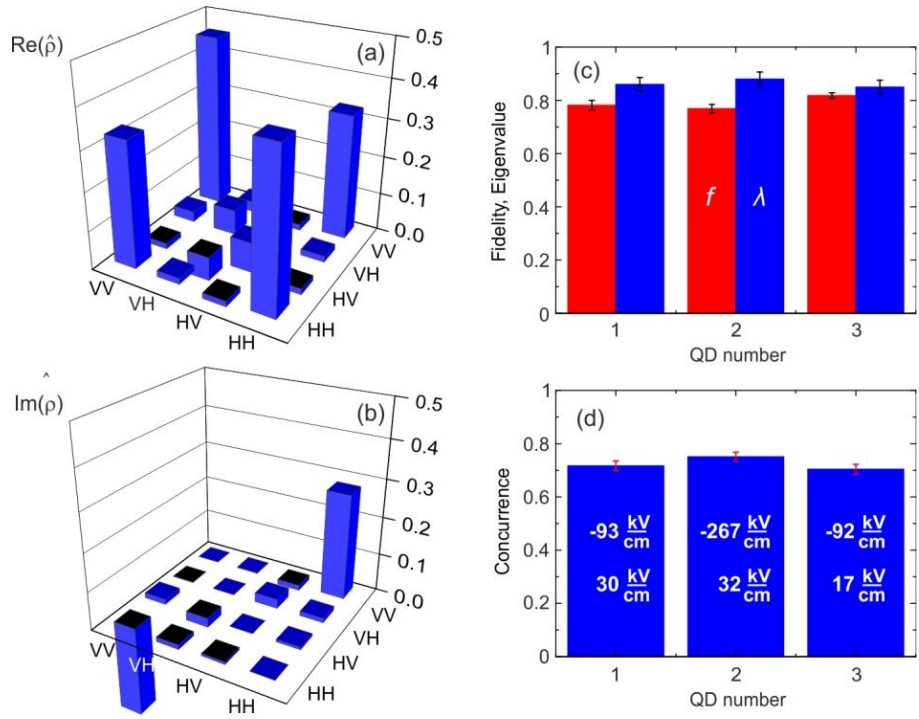

Figure 3 of 4
by R. Trotta *et al*.

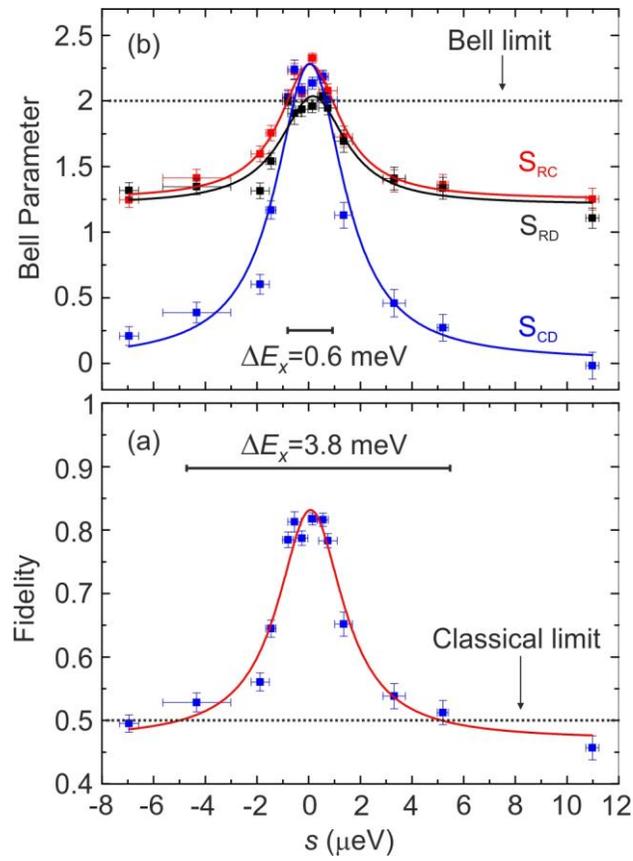

Figure 4 of 4
by R. Trotta *et al*.